\documentclass[12pt]{revtex4}
\usepackage{epsfig}
\usepackage{graphicx}
\newcommand{\eref}[1]{(\ref{#1})}
\newcommand{\na}{\mbox{\boldmath $\nabla$}}

\newcommand{\T}{\mbox{\boldmath $T$}}

\newcommand{\av}{\mbox{\boldmath $a$}}

\renewcommand{\j}{\mbox{\boldmath $j$}}

\newcommand{\ka}{\mbox{\boldmath $\kappa$}}
\newcommand{\ps}{\mbox{\boldmath $\psi$}}

\newcommand{\e}{\mbox{\boldmath $e$}}

\newcommand{\n}{\mbox{\boldmath $n$}}
\newcommand{\uu}{\mbox{\boldmath $u$}}

\newcommand{\ta}{\mbox{\boldmath $\tau$}}

\newcommand{\rr}{\mbox{\boldmath $r$}}

\newcommand{\J}{\mbox{\boldmath $J$}}

\newcommand{\B}{\mbox{\boldmath $B$}}

\newcommand{\kk}{\mbox{\boldmath $k$}}
\newcommand{\tk}{\mbox{\boldmath $\tilde{k}$}}
\newcommand{\skk}{\mbox{\scriptsize\boldmath $k$}}

\newcommand{\A}{\mbox{\boldmath$A$}}

\newcommand\fr{\displaystyle\frac}

\newcommand{\htts}{\mbox{\boldmath$\hat{t}\kern1pt$}}

\renewcommand{\tan}{{\rm\ tg}}

\newcommand\lt{\left}
\newcommand\rt{\right}

\newcommand{\srr}{\mbox{\scriptsize\boldmath$r$}}

\begin{document}
\titlepage
\begin{center}

{\Large\bf Those wonderful elastic waves}\\
\medskip

{\bf V.K.Ignatovich, L.T.N. Phan}\\
Joint Institute for Nuclear Research\\
Neutron Physics Laboratory
\bigskip

{\bf Abstract}

\end{center}

We consider in a simple and general way elastic waves in isotropic
and anisotropic media, their polarization, speeds, reflection from
interfaces with mode conversion, and surface waves. Reflection of
quasi transverse waves in anisotropic media from a free surface is
shown to be characterized by three critical angles.
\bigskip

\section{Introduction}

In our time of supercollider, quantum computing, teleportation,
dark matter and an eager search for a new physics, acoustics and
elastic waves look like an archaic science. Am.J.Phys. rarely
publish paper on this topic. In fact, we have found the single
article of 1980~\cite{leu} relevant to our consideration. Indeed
this science looks archaic, because everything seems to be well
resolved and the science became like an engineering tool,
frequently used in many applications. We will show that this
impression is wrong.

The theory seems to be well established (see, for
instance~\cite{land}), and all the textbooks~\cite{fesh}, are
unanimous in its presentation. The main notion is a displacement
vector $\uu(\rr,t)$ of a material point at a position $\rr$ at a
time moment $t$. Its Descartes components $u_i(\rr,t)$ obey the
Newtonian equation of motion
\begin{equation}\label{in}
\rho\fr{\partial^2}{\partial t^2}u_i(\rr,t)=\fr{\partial}{\partial
x_j}\sigma_{ij}(\rr,t),
\end{equation}
where $\rho$ is the material density, $x_j$ are components of the
radius-vector $\rr$, $\sigma_{ij}$ is a stress tensor
\begin{equation}\label{in2}
\sigma_{ij}=c_{ijkl}u_{kl},
\end{equation}
which is proportional to the deformation tensor $u_{ij}$
\begin{equation}\label{in1}
u_{jk}=\fr12\left(\fr{\partial u_j}{\partial x_k}+\fr{\partial
u_k}{\partial x_j}\right),
\end{equation}
and coefficients of proportionality $c_{ijkl}$ in \eref{in2}
comprise themselves a tensor with the symmetries:
\begin{equation}\label{in3}
c_{ijkl}=c_{jikl}=c_{ijlk}=c_{klij}.
\end{equation}
As is usual, in \eref{in}, \eref{in2} and everywhere below a
summation over repeated indices is assumed.

In isotropic media the tensor $c_{ijkl}$ is very simple:
\begin{equation}\label{in4}
c_{ijkl}=\lambda\delta_{ij}\delta_{kl}+\mu(\delta_{il}\delta_{kj}+\delta_{ik}\delta_{lj}),
\end{equation}
where $\delta_{ij}$ is the Kronecker symbol equal to unity for
$i=j$ and to zero otherwise, and $\lambda$, $\mu$ are two
parameters called Lam\'e elastic constants.

In the case of anisotropic media (usually crystals) the tensor
$c_{ijkl}$ contains considerably more
parameters~\cite{die,tru,mus,fed}. They are phenomenological, and
their physical meaning is not sufficiently clear.

The displacement vector $\uu(\rr,t)$ is usually represented as a
sum $\uu(\rr,t)=\na\varphi+\na\times\ps$ of two parts, where
$\varphi$ is a scalar, $\ps$ is a vector potentials, $\na$ is the
differential vector
\begin{equation}\label{in4n}
\na=\e_x\nabla_x+\e_y\nabla_y+\e_z\nabla_z\equiv\e_x\fr{\partial}{\partial
x}+\e_y\fr{\partial}{\partial y}+\e_z\fr{\partial}{\partial z},
\end{equation}
and $\e_{x,y,z}$ are unit vectors along three axes of an
orthogonal Descartes reference frame.

We will show that the use of the scalar and vector potentials is
not necessary. They only complicate the theory. Everything can be
presented in much more simple and transparent form with a wave
function, like in particle physics. In the case of isotropic media
such a presentation makes the theory looking almost trivial.

In the case of anisotropic media it is quite instructive to
consider not crystals, but a medium with an anisotropy
distinguished only by a single direction~\cite{nik}, because a
single vector is sufficient to elucidate the difference of
isotropic and anisotropic media. We found very interesting counter
intuitive features of waves reflected from interfaces in
anisotropic media. Surprisingly, but these features have not yet
been discussed in literature.

After general introduction in the second section to the theory of
elastic waves in isotropic and anisotropic media, we in the third
section consider isotropic ones. In isotropic media elastic waves
are very naturally break up into three classes, which are called
modes: two modes with transverse and one mode with longitudinal
polarizations.

Theory of waves in a homogeneous medium is rather primitive. It
becomes more rich when the medium contains an interface or a free
surface. In that case a wave reflects, refracts and multiply
splits at the interface. The presence of a surface or an interface
gives rise also to waves of the fourth mode --- the surface waves,
which run along the surface, exponentially decay away from it, and
have a mixed polarization. We consider how do they appear and what
are their properties.

In the fourth section we go to anisotropic media with a single
anisotropy vector. There again all waves break up into three
classes-modes, but only one mode has purely transverse
polarization. The other two are hybrids which are nor transverse
nor longitudinal. One of these modes is called quasi transverse
and the other one --- quasi longitudinal, because the smaller is
the anisotropy parameter, the closer is their polarization to pure
transverse and longitudinal directions respectively.

This is more or less evident and simple. The complications start
when a wave meets an interface or a free surface. In that case we
have reflection and refraction, which are in general accompanied
with triple waves splitting. Specular reflection is absent, and
surface waves become exotic. An experience acquired from isotropic
media considerations leads to a conclusion that in anisotropic
ones an incident wave can completely transform into a surface
wave! It is absolutely unacceptable because of the energy
conservation law. The incident wave carries an energy, which must
accumulate in the surface one, therefore the amplitude of the last
one must grow exponentially. It is impossible to describe such a
process by a linear stationary wave theory. So we meet a paradox:
the theory predicts some non physical solutions, which cannot be
described by the theory.

We considered the problem in details, and found that the reflected
waves do not accumulate into a surface mode because of unexpected
counter intuitive properties of the elastic waves in anisotropic
media. For instance, besides the first critical grazing angle
$\varphi_c$, similar to that one in isotropic media, at which a
quasi longitudinal wave becomes of the surface type, there is a
second critical angle, $\varphi_{c1}<\varphi_c$, when the
reflected wave, which is naturally thought of as moving away from
the reflecting surface, changes its direction, as if starting to
move toward it, though its energy flux remains going away from the
surface.

There is also a third critical angle, $\varphi_{c2}<\varphi_{c1}$,
at which the energy flux toward the surface becomes zero. This
angle is not zero, i.e. $\varphi_{c2}>0$, and annulation of energy
flux at this angle means that we cannot direct a ray of elastic
waves to the surface at an angle $\varphi<\varphi_{c2}$. It is
really strange, but we consider it as a hint, which the linear
theory of elasticity gives us to point out the cases, where we
have to involve a nonlinearity. We could not understand how to
introduce it, but we were so much impressed by the unexpected
properties of the elastic waves, that we decided to relate about
them to the readers of this journal.

Our research was started because of a need to explain the
difference between theoretically predicted and experimentally
measured anisotropy of sound speed in rocks ~\cite{nik}. The
theory, based on texture of rocks measured at a neutron
diffractometer, predicted anisotropy of speeds in some rocks three
times lower than the one measured with ultrasound. We hope that
our results will shed light on this difference, and will be
checked in an experiment of the kind discussed in~\cite{leu}.

\section{The main equations for elastic waves}

The starting point for study of elasticity is the free energy
density of a medium deformation~\cite{land}. For isotropic media
it is
\begin{equation}
\label{eq0} F = \frac{\lambda}{2}u_{ll}^2 + \mu u_{lj}^2,
\end{equation}
where $u_{ij}$ is the deformation tensor (\ref{in1}), $\lambda$,
$\mu$ are the Lam\'e elastic constants, notation $u_{lj}^2$ means
$u_{lj}u_{jl}$, and summation over repeated indices is assumed.

In anisotropic case we have to distinguish a direction, say along
a unit vector $\av$, and introduce a new elastic constant, say
$\zeta$, which is of the same dimension of energy density as
$\lambda$ and $\mu$. Then the free energy becomes
\begin{equation}
\label{eqq1} F = \frac{\lambda}{2}u_{ll}^2 + \mu u_{lj}^2 - \zeta
[(a_j u_{jl} )^2+(u_{jl}a_l)^2],
\end{equation}
where $a_j$ are Descart components of the vector $\av$,
anisotropic part is proportional to square of the vector with
components $a_ju_{ji}$, and the sign before $\zeta$ is not
necessary negative. Though the two anisotropic terms in square
brackets are identical ($u_{lj}$ is symmetrical) we put them
separately to get the tensor $c_{ijkl}$ with required symmetry
\eref{in3}.

The anisotropy term shows that the energy of deformation depends
not only on change of volume (the term $\propto\lambda$) and
change of shape (the term $\propto\mu$), but also on angles of
both deformations with respect to the anisotropy vector $\av$ (the
term $\propto\zeta$).

A question can be raised here: why do we use such an anisotropy
modification of the isotropic free energy \eref{eq0}? Is it not
possible to find a different one? The reply is yes. It is possible
to use a different modification. For instance, instead of
\eref{eqq1} we can accept the free energy in the form
\begin{equation}
\label{eeqq1} F = \frac{\lambda}{2}u_{ll}^2 + \mu u_{lj}^2 - \zeta
(a_j u_{jl}a_l)^2,
\end{equation}
or we can use a combinations of anisotropic terms in \eref{eeqq1}
and in \eref{eqq1}. We chose \eref{eqq1}, and it is an
arbitrariness. We think that consideration of different types of
anisotropy is a good task for students.

With the free energy we can define the stress tensor
\begin{equation}\label{eq2}
\sigma _{ij} = \fr{\partial F}{\partial u_{ij}},
\end{equation}
which after substitution of (\ref{eqq1}) gives
\begin{equation}
\label{eq4}\sigma _{ij} = \lambda \delta _{ij} u_{ll} + 2\mu
u_{ij} - 2\zeta (a_i u_{jl} a_l + a_l u_{li} a_j )=c_{ijkl}u_{kl},
\end{equation}
where
\begin{equation}\label{eq4a}
c_{ijkl}=\lambda\delta_{ij}\delta_{kl}+\mu(\delta_{il}\delta_{kj}+\delta_{ik}\delta_{lj})-\zeta
(a_i\delta_{jl}a_k+\delta_{il}a_ja_k+a_i\delta_{jk}a_l+\delta_{ik}a_ja_l).
\end{equation}
The expression for isotropic medium is obtained in the limit
$\zeta\to0$. Note that the tensor $c_{ijkl}$ satisfies the
symmetry requirements (\ref{in3}). In the case of \eref{eeqq1} the
anisotropic part of the tensor $c_{ijkl}$ will be simply a product
$\zeta a_ia_ja_ka_l$.

With the stress tensor (\ref{eq4}) the Newtonian equation of
motion (\ref{in}) for the displacement vector becomes
$$\rho \ddot {u}_i = \nabla _j \sigma_{ij} = \mu [\Delta u_i
+ \nabla _i (\nabla\cdot\uu)] + \lambda \nabla _i (\na\cdot\uu)-$$
\begin{equation}
\label{eq5}
 - \zeta \left(a_i [\Delta (\uu\cdot\av) + (\av\cdot\na)(\na\cdot\uu)] + (\av\cdot\na)^2u_i + \nabla _i
(\av\cdot\na)(\av\cdot\uu)\right),
\end{equation}
where $\rho$ is the medium density.

We can seek solution of (\ref{eq5}) in the form of a complex plain
wave $\uu(\rr,t)=\A\exp(i\kk\rr-i\omega t)$, like a wave function
in particle physics, where vector $\A$ ia a unit polarization
vector. Of course, the elastic waves are real waves, so they are
represented by the real part of the complex wave function. Later
we shall see, where we must be especially careful in description
of elastic waves with complex function, but for now we meet no
difficulties, and after substitution of such $\uu$ into
(\ref{eq5}) we obtain an equation for $\A$:
$$\rho \omega ^2\A = \mu k^2\A + (\lambda + \mu
)\kk(\kk\cdot\A)-$$
\begin{equation}
\label{eeq6} - \zeta \left( \av[k^2(\av\cdot\A) +
(\kk\cdot\av)(\kk\cdot\A)] + (\kk\cdot\av)[(\kk\cdot\av)\A
+\kk(\av\cdot\A)] \right).
\end{equation}
It is convenient to transform this equation to dimensionless form
dividing both parts of (\ref{eeq6}) by $\mu k^2$. After
introduction of the standard transverse speed
$c_t=\sqrt{\mu/\rho}$, the phase speed of the wave $V=\omega/k$,
dimensionless ratio $v=V/c_t$, dimensionless parameters
$E=(\lambda+\mu)/\mu$, $\xi=\zeta/\mu$ and the unit vector
$\ka=\kk/k$ the equation becomes
\begin{equation}
\label{eq7} \Omega^2\A = E\ka(\ka\cdot\A)- \xi\Big[\{
\av(\av\cdot\A) +
(\ka\cdot\av)\Big[\av(\ka\cdot\A)+\A(\av\cdot\ka)
+\ka(\A\cdot\av)\Big] \Big\},
\end{equation}
where $\Omega^2=v^2-1$.
First we put $\xi=0$, and consider an
isotropic medium.

\section{Waves in isotropic media ($\xi=0$)}

For isotropic media the equations (\ref{eq5}-\ref{eq7}) are
reduced respectively to
\begin{equation}
\label{eq5a}
 \rho \ddot {u}_i = \nabla _j u_{ij} = \mu [\Delta u_i
+ \nabla _i (\nabla\cdot\uu)] + \lambda \nabla _i
(\na\cdot\uu),\end{equation}
\begin{equation}
\label{eq6a} \rho \omega ^2\A = \mu k^2\A + (\lambda + \mu
)\kk(\kk\cdot\A),
\end{equation}
\begin{equation}
\label{eq7a} \Omega^2\A = E\ka(\ka\cdot\A).
\end{equation}
For the given propagation direction $\ka$ we can introduce two
orthonormal vectors $\e^{(1)}$ and $\e^{(2)}$, which are
perpendicular to $\ka$. In the orthonormal basis $\e^{(1)}$,
$\e^{(2)}$, $\ka$ the polarization unit vector $\A$ is
representable as
\begin{equation}\label{eq8}
\A=\alpha^{(1)}\e^{(1)}+\alpha^{(2)}\e^{(2)}+\beta\ka.
\end{equation}
After multiplication of (\ref{eq7a}) consecutively by $\e^{(1,2)}$
and $\ka$ we obtain three equations for coordinates
$\alpha^{(1,2)}$ and $\beta$:
\begin{equation}\label{eq9}
\Omega^2\alpha^{(1,2)}=0,\qquad [\Omega^2-E]\beta=0.
\end{equation}
They are independent and give three solutions
\begin{equation}\label{eeq9}
\A^{(1,2)}=\e^{(1,2)},\qquad \A^{(3)}=\ka.
\end{equation}
Their speeds are determined by equations
\begin{equation}\label{eeq9a}
(v^{(1,2)})^2=1\to V^{(1,2)}=\sqrt{\fr\mu\rho}=c_t,\qquad
(v^{(3)})^2=E+1\to
V^{(3)}=\sqrt{E+1}c_t=\sqrt{\fr{\lambda+2\mu}{\rho}}=c_l.
\end{equation}
Since $V=\omega/k$, we can tell that for a given frequency
$\omega$ the wave numbers $k=|\kk|$ of three modes are
\begin{equation}\label{eqq9}
k^{(1,2)}=\fr\omega{c_t},\qquad k^{(3)}=\fr\omega{c_l}.
\end{equation}

The two unit vectors $\e^{(1,2)}$ above are orthogonal to each
other, and lie in the plane, perpendicular to the unit vector
$\ka$, but their azimuthal angle around $\ka$ can be arbitrary. We
can use this freedom to facilitate solution of different problems.
In particular, below, when we consider reflection from an
interface. There we can choose $\e^{(1)}$ to be perpendicular to
the incidence plane, and $\e^{(2)}$ to be inside it.

\subsection{Reflection from an interface}

Suppose that the medium consists of two parts: one at $z<0$ with
constants $\lambda$, $\mu$, $\rho$ and another one at $z>0$ with
constants $\lambda'$, $\mu'$, $\rho'$, then  there is reflection
and refraction of waves at the interface $z=0$. If a plane wave
$\uu_{inc}(\rr,t)=\A\exp(i\kk\rr-i\omega t)$ incident from $z<0$
is of mode $\A^{(j)}$ ($j$ is one of the numbers 1, 2 or 3) then
the interface transforms this displacement vector to

$$\uu(\rr,t)=\exp(i\kk_\|\rr_\|-i\omega
t)\times$$
\begin{equation}\label{eq10}
\times\left[\left(\A^{(j)}e^{ik^{(j)}_\bot
z}+\sum\limits_{l=1}^3r^{(jl)}\A^{(l)}_Re^{-ik^{(l)}_{\bot}z}\right)\Theta(z<0)+
\sum\limits_{l=1}^3t^{(jl)}\A^{(l)}_{T}e^{ik'^{(l)}_{\bot}z}\Theta(z>0)\right],
\end{equation}
where $r^{(jl)}$, $t^{(jl)}$ are reflection and refraction
amplitudes of the $l$-th mode ($l=1,2,3$) for the incident $j$-th
mode, and $\Theta$ is a step function, which is equal to unity,
when inequality in its argument is satisfied, and to zero
otherwise.

Below the interface ($z<0$) the displacement consists of the
incident wave of $j$-th mode and reflected waves of modes
$\A^{(l)}_{R}$ (the lower index $R$ means reflected). Above the
interface ($z>0$) the displacement consists of transmitted waves
of modes $\A^{(l)}_{T}$ (the lower index $T$ means transmitted).

\begin{figure}[!b]
{\par\centering\resizebox*{10cm}{!}{\includegraphics{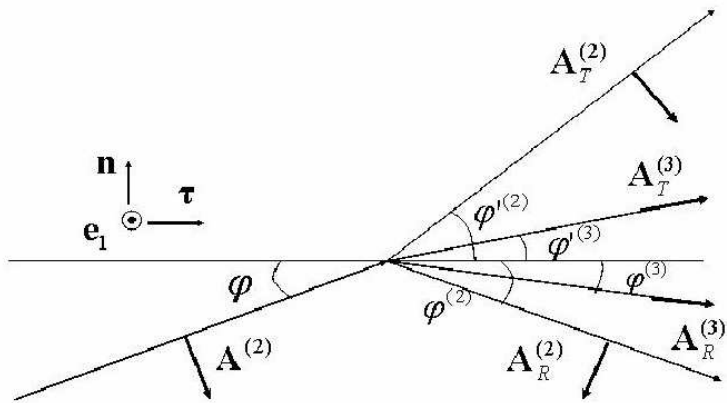}}
\par}
\caption{\label{f1}Reflection of a transverse wave $\A^{(2)}$ from
an interface between two different isotropic media. Reflected and
refracted waves contain two modes: $\A^{(2,3)}_R$ and
$\A^{(2,3)}_T$ respectively. The $\A^{(2)}_R$ mode goes at
specular grazing angle $\varphi^{(2)}=\varphi$, the longitudinal
mode $\A^{(3)}_R$ goes at grazing angle
$\varphi^{(3)}<\varphi^{(2)}$.}
\end{figure}

All the waves differ from one another not only by polarization,
but also by the wave vector $\kk$, which can be represented as
\begin{equation}\label{eqq10}
\kk=\kk_\|+\kk_\bot\equiv\ta k_\|\pm\n k_\bot,
\end{equation}
where $\ta$ is a unit vector along the interface, and $\n$ is a
unit vector along $z$-axis perpendicular to the interface, as is
shown in Fig. \ref{f1}. Note, that the component, $\kk_\|=\ta
k_\|$, of the wave vectors is identical for all the waves, as is
demonstrated by the first common factor
$\exp(i\kk_\|\rr_\|-i\omega t)$ in \eref{eq10}, where $\rr_\|$ are
coordinates in the interface. The vector $\kk_\|$ is identical,
because the space along $\ta$ is uniform and nothing can change
this component.

The normal components $k^{(l)}_{\bot}$ and $k'^{(l)}_{\bot}$ of
wave vectors of modes $\A^{(l)}_R$ and $\A^{(l)}_T$ respectively,
are positive numbers and their value depends on the mode $l$.
Since $k_\bot^2=k^2-k_\|^2$, then from (\ref{eqq9}) it follows
that
\begin{equation}\label{11}
k^{(1,2)}_{\bot}=\sqrt{\fr{\omega^2}{c_{t}^2}-k_\|^2},\quad
k'^{(1,2)}_{\bot}=\sqrt{\fr{\omega^2}{c'^2_t}-k_\|^2},\quad
k^{(3)}_{\bot}=\sqrt{\fr{\omega^2}{c_{l}^2}-k_\|^2},\quad
k'^{(3)}_{\bot}=\sqrt{\fr{\omega^2}{c'^2_{l}}-k_\|^2},
\end{equation}
where $c_{t,l}$, and $c'_{t,l}$, are the speeds defined in
\eref{eeq9a} for lower and upper spaces respectively.

To find reflection and refraction amplitudes, we need boundary
conditions. One of them is continuity of the displacement vector:
\begin{equation}\label{eq12}
\uu|_{z=-0}=\uu|_{z=+0}\to\A^{(j)}+\sum\limits_{l=1}^3r^{(jl)}\A^{(l)}_{R}=
\sum\limits_{l=1}^3t^{(jl)}\A^{(l)}_{T},
\end{equation}
and the second one is the continuity of the stress vector $\T$
with components $T_j=\sigma_{jl}n_l$. According to (\ref{eq4})
this vector for a displacement $\uu(\rr,t)$ is equal to
\begin{equation}
\label{eq13}\T\Big(\uu(\rr,t)\Big)=\lambda
\n(\na\cdot\uu)+\mu[\na(\uu\cdot\n)+(\n\cdot\na)\uu].
\end{equation}
Continuity of the vector $\T$ is equivalent to the equation
\begin{equation}\label{eq12a}
\B^{(j)}+\sum\limits_{l=1}^3r^{(jl)}\B^{(l)}_{R}=
\sum\limits_{l=1}^3t^{(jl)}\B^{(l)}_{T},
\end{equation}
where the vector $\B$ is defined as
\begin{equation}\label{eq12b}
\B=-i\exp(-i\kk\rr)\T\Big(\A\exp(i\kk\rr)\Big)=\lambda
\n(\kk\cdot\A)+\mu[\kk(\A\cdot\n)+\A(\n\cdot\kk)],
\end{equation}
for every plane wave $\A\exp(i\kk\rr)$.

Note that the condition (\ref{eq12a}) makes it possible to
continue the wave equation (\ref{eq5}) from $z<0$ to $z>0$. If it
is not satisfied, the differentiation of $\sigma_{ij}$ in
(\ref{eq5}) creates $\delta(z)$-function and the wave equation
becomes inhomogeneous~\cite{ign} with a source term at the
interface $z=0$.

To find reflection and transmission amplitudes we need to multiply
both equations (\ref{eq12}) and (\ref{eq12a}) by three mutually
orthogonal unit vectors to get in general 6 equations for 6
unknowns. It is convenient to choose the right triple the vectors
$\ta$, $\e^{(1)}$, $\n$, as shown in Fig,~\ref{f1}, where the
vector $\e^{(1)}$ is perpendicular to the incidence plane and in
Fig.~\ref{f1} points toward the reader.

Before calculations of the amplitudes of all the reflected and
refracted waves we can easily understand what are all the angles.
The grazing angle $\varphi$ of the wave with the wave vector $\kk$
is defined via relation
$\cos\varphi=\ta\cdot\kk/k=k_\|/k=k_\|V/\omega$. Since $k_\|$ and
$\omega$ are identical for all the waves therefore the value
$\cos\varphi^{(j)}/V^{(j)}$ are also the same for all the waves.
And because of \eref{eeq9a} we can write
\begin{equation}\label{fi23}
\fr{\cos\varphi}{V^{(j)}}=\fr{\cos\varphi^{(1,2)}}{c_t}=\fr{\cos\varphi^{(3)}}{c_l}=
\fr{\cos\varphi'^{(1,2)}}{c'_t}= \fr{\cos\varphi'^{(3)}}{c'_l},
\end{equation}
where $\varphi^{(i)}$ $\varphi'^{(i)}$ denote grazing angle for
respectively reflected and transmitted waves of mode $i$, and
$\varphi$ without indices denotes the grazing angle of the
incident wave.

To find directions of polarization after reflection is very easy
and we leave it as an exercise for the reader to check that with
account of \eref{eqq9}
\begin{equation}\label{tri}
\A^{(2)}_{R}=[\ka^{(2)}_{R}\times\e^{(1)}]=-\fr{k^{(2)}_{\bot}\ta+k_\|\n}{k^{(2)}},
\quad \A^{(3)}_{R}=\fr{-k^{(3)}_{\bot}\n+k_\|\ta}{k^{(3)}}.
\end{equation}

\subsubsection{Reflection of $\A^{(1)}$ mode}

The simplest is reflection and refraction of $\A^{(1)}$ mode. Its
polarization is $\e^{(1)}$. After multiplication of equations
(\ref{eq12}) and (\ref{eq12a}) by $\e^{(1)}$ we get
\begin{equation}\label{15}
1+r^{(11)}= t^{(11)}, \qquad
\mu_1(1-r^{(11)})k_\bot=\mu_2t^{(11)}k'_\bot,
\end{equation}
from which it immediately follows that
\begin{equation}\label{13}
r^{(11)}=\fr{\mu k_\bot-\mu'k'_\bot}{\mu k_\bot+\mu'k'_\bot},
\end{equation}
where $k_\bot=\sqrt{\omega^2/c^2_t-k_\|^2}$, and
$k'_\bot=\sqrt{\omega^2/c'^2_t-k_\|^2}$. We see that this mode is
reflected specularly and no other modes are created.

\subsubsection{Reflection of $\A^{(2)}$ mode, and the mode conversion.}

The more interesting is the case of the incident $\A^{(2)}$ mode
shown in Fig. \ref{f1}. Its reflection and refraction creates
longitudinal mode $\A^{(3)}_{R}$, and because the speed $c_l$ of
$\A^{(3)}_{R}$ is larger than the speed $c_t$ of the specularly
reflected $\A^{(2)}_{R}$ mode, the grazing angle $\varphi^{(3)}$
is less than $\varphi^{(2)}$. Therefore we can expect that at some
angle $\varphi=\varphi_c$ of the incident wave, the angle
$\varphi^{(3)}$ becomes zero, which means that the longitudinal
mode ceases to propagate in the direction $z<0$. Since according
to \eref{fi23} $\cos\varphi^{(3)}=(c_l/c_t)\cos\varphi\le1$, we
find that $\varphi_c=\arccos(c_t/c_l)$. The similar considerations
are applicable to the refracted waves, and we can expect that at
$\varphi<\arccos(\max(c_t/c_l,c'_t/c'_l)$ there appears a purely
longitudinal surface wave propagating along the interface.

To find amplitudes of the reflected and refracted modes we have to
multiply the two equations, (\ref{eq12}) and (\ref{eq12a}), by
$\n$ and $\ta$. As a result we get a linear system of four
equations for 4 unknown $r^{(22)}$, $r^{(23)}$, $t^{(22)}$ and
$t^{(23)}$, which can be solved analytically. However it is a
boring job, so it is better to pass it to computer.

The analytical solution can be found for reflection from a free
surface, where we have a single boundary condition
\begin{equation}\label{eq12d}
\B^{(2)}+r^{(22)}\B^{(2)}_{R}+r^{(23)}\B^{(3)}_{R}=0.
\end{equation}
In this case we have only two reflected waves and multiplication
of (\ref{eq12d}) by $\n$ and $\ta$ with account of (\ref{eq12b})
and (\ref{tri}) gives only two equations
\begin{equation}\label{q4}
-\fr{2k^{(2)}_{\bot}k_\|}{k^{(2)}}(1-r^{(22)})+r^{(23)}\fr{k^{(2)2}-
k_{\|}^2}{k^{(3)}}=0,
\end{equation}
\begin{equation}\label{q4a}
\fr{k^{(2)2}-2k_\|^2}{k^{(2)}}(1+r^{(22)})-2
r^{(23)}\fr{k^{(3)}_{\bot}k_\|}{k^{(3)}}=0.
\end{equation}
Their solution is
\begin{equation}\label{q5}
r^{(23)}=\fr{k^{(3)}}{k^{(2)}}\,
\fr{4k^{(2)}_{\bot}k_\|(k^{(2)2}-2k_\|^2)}{4k^{(3)}_{\bot}k^{(2)}_{\bot}k_\|^2+(k^{(2)2}-2k_{\|}^2)^2},\qquad
r^{(22)}=\fr{4k^{(3)}_{\bot}k^{(2)}_{\bot}k_\|^2-(k^{(2)2}-2k_{\|}^2)^2}{4k^{(3)}_{\bot}k^{(2)}_{\bot}k_\|^2+(k^{(2)2}-2k_{\|}^2)^2}.
\end{equation}

At $\varphi=\varphi_c$ the vector $\ka^{(3)}$ of the longitudinal
wave propagation direction coincides with $\ta$, and therefore the
length $k^{(3)}=\omega/c_l$ of the wave vector $\kk^{(3)}$ becomes
equal to $k_\|$. When the grazing angle $\varphi$ decreases below
$\varphi_c$ the value of $k_\|$ increases, but $\omega/c_l$ does
not change. Therefore at $\varphi<\varphi_c$ we get
$k_\|>\omega/c_l$, and
$k^{(3)}_{\bot}=\sqrt{\omega^2/c_l^2-k_\|^2}$ becomes imaginary.
We can denote it $-iK_l$. With such a normal component of the wave
vector the longitudinal wave, propagating along the free surface,
becomes localized in the layer of thickness $l=1/K_l$, where
$K_l=\sqrt{k_\|^2-\omega^2/c_l^2}$. In other words, it becomes
longitudinal surface wave $\uu^{(3)}_{S}$ (lower index $S$ means
surface) with complex polarization vector $\A^{(3)}_{S}$. But what
a strange wave it is! Since the incident wave can have arbitrary
$\omega$ and $k_\|<k=\omega/c_t$, the longitudinal surface wave
with the same $\omega$ and $k_\|$, has the speed along the
surface, (denote it $V^{(3)}_{S}$) equal to
$V^{(3)}_{S}=\omega/k_\|>c_t$, which means, that it is not the
Rayleigh surface wave, because the speed $c_R$ of the Rayleigh
wave, as is well known, is less than $c_t$!

However, really, it is not strange. This longitudinal surface wave
satisfies the same wave equation $\Omega^2=E$ of (\ref{eq9}), and
has longitudinal polarization
\begin{equation}\label{ls}
\A^{(3)}_{S}=(k_\|\ta-iK_l\n)/\sqrt{k_\|^2+K_l^2},
\end{equation}
i.e. its normal component is imaginary. It is not dangerous that
this polarization is a complex vector. The displacement must have
a real value, therefore the displacement with a complex
polarization vector is
\begin{equation}\label{ls1}
\uu^{(3)}_{S}\propto{\rm
Re}\left[(k_\|\ta-iK_l\n)e^{i\skk_\|\srr_\|+K_l z-i\omega
t}\right]=\left[(k_\|\ta\cos(\kk_\|\rr_\|-\omega
t)+K_l\n\sin(\kk_\|\rr_\|-\omega t)\right]e^{K_l z},
\end{equation}
i.e. the phase of oscillations along vector $\n$ is shifted by
$\pi/2$ with respect to oscillations along vector $\ta$. The speed
of the longitudinal wave along the interface,
$V^{(3)}_{S}=\omega/k_\|$, can be arbitrary, though because of
$k_\|^2>\omega^2/c_l^2$, this speed lies in the interval
$c_t<V^{(3)}_{S}<c_l$.

In a similar way we can define the transverse surface wave. It
satisfies the wave equation $\Omega^2=0$ of (\ref{eq9}), and has
transverse complex polarization
\begin{equation}\label{ts}
\A^{(2)}_{S}=(k_\|\n+iK_t\ta)/\sqrt{k_\|^2+K_t^2},
\end{equation}
where $K_t=\sqrt{k_\|^2-\omega^2/c_t^2}$. The real displacement
vector in it is
\begin{equation}\label{ts1}
\uu^{(2)}_{S}\propto{\rm
Re}\left[(k_\|\n+iK_t\ta)e^{i\skk_\|\srr_\|+K_t z-i\omega
t}\right]=\left[(k_\|\n\cos(\kk_\|\rr_\|-\omega
t)-K_t\ta\sin(\kk_\|\rr_\|-\omega t)\right]e^{K_t z},
\end{equation}
and the speed $V^{(2)}_{S}$ along the interface can be arbitrary
but less than $c_t$.

When longitudinal wave is of the surface type, the reflection
amplitude of the $\A^{(2)}_{R}$ mode according to \eref{q5}
becomes
\begin{equation}\label{e8}
r^{(22)}=-\fr{(k^{(2)2}-2k_{\|}^2)^2-4iK_lk^{(2)}_{\bot}k_\|^2}{(k^{(2)2}-2k_{\|}^2)^2+4iK_lk^{(2)}_{\bot}k_\|^2}.
\end{equation}
It is a unit complex number, therefore it describes the total
reflection of the incident wave.

\subsubsection{Energy flux distribution between two reflected
waves}

Because of energy conservation the energy flux density of the
incident wave along the normal to the interface must be equal to
the sum of energy flux densities of the reflected waves. Let's
check, whether they are really equal.

When the displacement and therefore the stress tensor are real
functions, the energy flux density of an elastic wave is described
by a vector $\j$ with components
\begin{equation}\label{jn}
j_i=-\langle\sigma_{il}du_l/dt\rangle,
\end{equation}
where $\langle F\rangle$ means averaging of the function $F$ over
time. We use displacement in the form of complex plane waves, but
the energy flux density should have only a real value, therefore
the Eq. \eref{jn} can be represented as~\cite{kis}
\begin{equation}\label{j1nn}
j_i=-\fr12\lt[\sigma^*_{il}\fr{du_l}{dt}+\sigma_{il}\fr{du^*_l}{dt}\rt]=-{\rm
Re}\lt[\sigma^*_{il}\fr{du_l}{dt}\rt],
\end{equation}
where Re($F$) means real part of $F$, and $*$ means complex
conjugation. We are interested in the flux density along the
normal $\n$ to the interface, therefore we need to calculate
$\n\j=\omega$Re$(i\T^*\uu)$, where we used \eref{eq13}. Taking
into account the definition \eref{eq12b} we can represent the
energy flux in the form
\begin{equation}\label{sff3}
\fr{(\j\cdot\n)}{\mu\omega}=\fr1\mu{\rm Re}(\B^*\cdot\A)=E{\rm
Re}((\A^*\cdot\n)(\kk\cdot\A)+ k_\bot).
\end{equation}

In the case of the incident $\A^{(2)}$ mode its incident flux is
$j^{(2)}=\mu\omega k^{(2)}_{\bot}=\rho\omega c_t^2k^{(2)}_{\bot}$.
Reflected fluxes of the two modes are $j^{(2)}_{R}=\rho
c_t^2|r^{(22)}|^2k^{(2)}_{\bot}$, $j^{(3)}_{R}=\rho
c_l^2|r^{(23)}|^2k^{(3)}_{\bot}$ Energy conservation law requires
\begin{equation}\label{sf4}
|r^{(22)}|^2+|r^{(23)}|^2\fr{c_l^2
k^{(3)}_{\bot}}{c_t^2k^{(2)}_{\bot}}=1.
\end{equation}
Substitution of (\ref{q5}) shows that this equation is satisfied.

When the longitudinal wave becomes of surface type, it does not
produce a flux from the surface. Therefore the law of energy
conservation (\ref{sf4}) reduces to
\begin{equation}\label{sf5}
|r^{(22)}|^2=1,
\end{equation}
which, according to (\ref{e8}), is also satisfied.

\subsection{Energy density of the longitudinal surface wave}

There is also one interesting question: what the energy density is
accumulated in the longitudinal surface wave. This question is
interesting, because it is this wave can be  important for
predictions and estimation of magnitudes of the earthquakes.

The vector of energy density flux of a wave of mode $\A$ according
to (45) can be represented as
\begin{equation}\label{enr}
\J={\rm Re}(\rho\omega c_t^2u_0^2[E\A^*(\kk^*\cdot\A)+\kk^*]),
\end{equation}
where we introduced an amplitude $u_0$ of the wave. Expression
\eref{enr} is valid for real and complex wave vectors and
polarizations. The absolute value of this flux for the incident
transverse wave is
\begin{equation}\label{enr1}
J=\rho\omega c_t^2u_0^2k.
\end{equation}
For the surface longitudinal wave with account of its factor
$r^{23}$ we get
\begin{equation}\label{enr2}
J_S^{3}(z)=\rho\omega u_0^2|r^{23}|^2c_l^2k_\|\exp(2K_lz).
\end{equation}
\begin{figure}[!th]
{\par\centering\resizebox*{10cm}{!}{\includegraphics{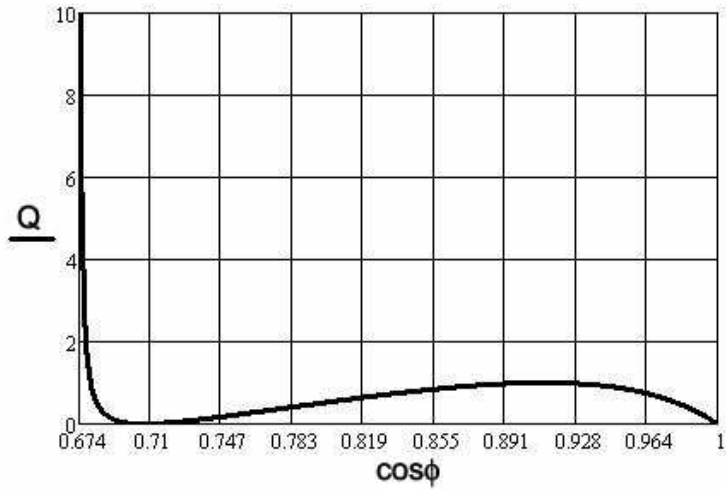}}\par}
\caption{\label{quak} Dependence of $Q$ on $\cos\varphi$
calculated for $E=1.2$. The left point on the abscise axis
corresponds to critical $\cos\varphi_c=0.674$, where $Q=323.16$.}
\end{figure}
We do not know $u_0$, so we can find only ratio
$Q=J_S^{3}(z=0)/J$. Substitution of $r^{23}$ from \eref{q5} gives
\begin{equation}\label{enr3}
Q=\fr{k_\|}{k^{(2)}}\fr{c_l^2}{c_t^2}|r^{23}|^2=\fr{k_\|}{k^{(2)}}
\fr{16k^{(2)2}_{\bot}k_\|^2(k^{(2)2}-2k_\|^2)^2}{16K_l^2k^{(2)2}_{\bot}k_\|^4+(k^{(2)2}-2k_{\|}^2)^4},
\end{equation}
or
\begin{equation}\label{enr4}
Q=
\fr{\cos\varphi\sin^2(4\varphi)}{4(\cos^2\varphi-\cos^2\varphi_c)\sin^2(2\varphi)+\cos^4(2\varphi)}.
\end{equation}
Dependence of this function on $\cos\varphi$ is shown in
Fig.~\ref{quak}. We see that the highest energy density is
accumulated in longitudinal surface wave, when $\varphi$ is
slightly less than $\varphi_c$. There is also a maximum at small
angles where the ratio $Q$ is close to unity.

\subsection{The Surface Rayleigh wave}

We considered above the two surface waves, which satisfy the wave
equations, but cannot exist independently, because without the
incident and reflected waves they  do not satisfy the boundary
condition. The Rayleigh surface wave exists without the incident
one, and its speed $c_R=\omega/k_\|<c_t$ is fixed. To get equation
which determines this speed $c_R$ we represent the boundary
condition (\ref{eq12d}) in the form
\begin{equation}\label{eq12d1}
\fr1{r^{(22)}}\B^{(2)}+\B^{(2)}_{R}+\fr{r^{(23)}}{r^{(22)}}\B^{(3)}_{R}=0,
\end{equation}
where $r^{(22)}$ and $r^{(23)}$ are given by (\ref{q5}). With
(\ref{eq12d1}) we can immediately find the speed of the Rayleigh
surface wave. It corresponds to such $\omega/k_\|$, for which the
first term in \eref{eq12d1} is zero. Since $\B^{(2)}\ne0$,
therefore the first term is zero only when
\begin{equation}\label{e11}
\fr1{r^{(22)}}=0.
\end{equation}
In such a case the incident wave disappears, and the whole wave
field contains only two waves propagating along the free surface.

Let's remind that a similar trick helps to find bound states of
particles in quantum mechanics. Reflection of a particle from a
one dimensional potential well is described in asymptotic region
$x\to -\infty$ by the wave function $\exp(ikx)+r(k)\exp(-ikx)$,
where $r(k)$ is a reflection amplitude. This wave function can be
also represented as $(1/r(k))\exp(ikx)+\exp(-ikx)$. In bound
states the wave function at $x\to -\infty$ has asymptotics
$\exp(-Kx)$, where $-K^2$ is proportional to the bound level
$E_b$. To find $K$ we need to solve equation $1/r(k)=0$, which
annuls the incident wave $\exp(ikx)$. Every root of this equation
$k_n=-iK_n$ corresponds to $n$-th bound level $E_{bn}\propto
-K_n^2$. In that respect the Rayleigh surface wave is a bound
state of elastic waves,

After this digression we go back. From (\ref{e8}) it follows that
(\ref{e11}) is satisfied, if
\begin{equation}\label{e12}
4k^{(3)}_{\bot}k^{(2)}_{\bot}k_\|^2+(k^{(2)2}-2k_{\|}^2)^2=0.
\end{equation}
It is important to note that the third term in \eref{eq12d1} does
not disappear though it also contains the factor $1/r^{(22)}$. It
does not disappear because $r^{(23)}$ and $r^{(22)}$ according to
\eref{q5} have the similar denominators, and they cancel each
other in the ratio $r^{(23)}/r^{(22)}$. In fact the amplitudes
$r^{(23)}$ and $r^{(22)}$ play equal roles, so instead of
\eref{eq12d1} we can write
\begin{equation}\label{eq12d1d}
\fr1{r^{(23)}}\B^{(2)}+\fr{r^{(22)}}{r^{(23)}}\B^{(2)}_{R}+\B^{(3)}_{R}=0,
\end{equation}
and seek solution of the equation $1/r^{(23)}=0$. The result will
be the same.

Let's denote the speed of the wave propagation, $\omega/k_\|$,
along the interface by $c_R$ (speed of the Rayleigh wave), and its
ratio to $c_t$ by $x=c_R/c_t$. Since $k^{(2)}_{\bot}$ and
$k^{(3)}_{\bot}$ in surface waves are to be imaginary then
$k^{(2)2}_{\bot}=\omega^2/c_t^2-k_\|^2<0$,
$k^{(3)2}_{\bot}=\omega^2/c_l^2-k_\|^2<0$, and the equation
(\ref{e12}) is reduced to
\begin{equation}\label{e14}
4\sqrt{1-x^2}\sqrt{1-\varsigma^2 x^2}=(2-x^2)^2,
\end{equation}
where $\varsigma=c_t/c_l$. This equation and its solution are well
known and can be found in all the textbooks on elastic waves. It
can be solved even analytically for arbitrary $\varsigma$, because
it is equivalent to an algebraic equation of 4-th order with
respect to variable $z=1-x^2$.

When (\ref{e14}) is satisfied, the total displacement vector of
the Rayleigh wave becomes
\begin{equation}\label{e14a}
\uu_R\propto\uu^{(2)}_{S}+\fr{r^{(23)}}{r^{(22)}}\uu^{(3)}_{S},
\end{equation}
Substitution of (\ref{q5}), (\ref{ls1}) and (\ref{ts1}) into
(\ref{e14a}) with the same $\omega/k_\|=c_R$ gives~\cite{vic}
\begin{equation}\label{rw}
\uu_R\propto\n\cos(\kk_\|\rr_\|-\omega
t)[2q_tq_le^{K_lz}-(1+q_t^2)e^{K_tz}]-\ta\sin(\kk_\|\rr_\|-\omega
t) q_t [2e^{K_lz}-(1+q_t^2)e^{K_tz}].
\end{equation}
where $K_{t,l}=k_\|\sqrt{1-c_R^2/c_{t,l}^2}$, and
$q_{t,l}=K_{t,l}/k_\|=\sqrt{1-c_R^2/c_{t,l}^2}$.

In a similar way we can find the Stoneley surface wave propagating
along the interface between two isotropic media. Though there are
no principal difficulties, we do not consider it here because of
technical complications.

From (\ref{eq12b}) we can immediately conclude that the surface
waves with polarization along the surface and perpendicular to
direction of propagation do not exist, because the continuity of
the stress vector $\B$ requires continuity of the normal
derivative of the displacement vector, which cannot be satisfied.

\section{Waves in anisotropic media ($\xi\ne0$)}

In isotropic media it was natural to describe polarizations in an
orthogonal basis $\e^{(1)}$, $\e^{(2)}$ and $\ka=\kk/k$, which
constitutes the right hand triple of unit vectors. The choice of
$\e^{(1)}$ and $\e^{(2)}$ had some freedom because these two
vectors can be rotated by an arbitrary angle around $\ka$. It was
only in studying of reflection from an interface, where
orientation of $\e^{(1)}$, $\e^{(2)}$ was fixed by the plane of
incidence. In anisotropic media besides $\ka$ we have also vector
$\av$, so for orientation of vectors $\e^{(1)}$, $\e^{(2)}$ it is
better to take the plane of vectors $\ka$ and $\av$  into account,
choosing $\e^{(2)}$ in the plane, and
$\e^{(1)}=[\e^{(2)}\times\ka]$ perpendicular to it.

After substitution of (\ref{eq8}) into (\ref{eq7}) and
multiplication by three vectors $\e^{(1)}$, $\e^{(2)}$ and
$\ka=\kk/k$ we obtain the system of linear equations for
$\alpha^{(1,2)}$, and $\beta$. It has a solution, if its
determinant is equal to zero. This condition gives three possible
possible speeds $V^{(i)}(k)$ (i=1,2,3) for the three wave modes.

The simplest equation is obtained after multiplication of the
(\ref{eq7}) by $\e^{(1)}$. The result is
\begin{equation}
\label{eq7a1} \Omega^2\alpha^{(1)} +
\xi(\ka\cdot\av)^2\alpha^{(1)}=0.
\end{equation}
This Eq. is equivalent to
\begin{equation}
\label{eq7b} \Omega^2 + \xi(\ka\cdot\av)^2=0,
\end{equation}
and it gives the speed of this transverse mode
\begin{equation}
\label{eq7c} V^{(1)}\equiv
\omega/k=c_t\sqrt{1-\xi(\ka\cdot\av)^2}=c_t\sqrt{1-\xi\cos^2\theta},
\end{equation}
where $\theta$ is the angle between vectors $\ka$ and $\av$. We
see that this speed is less than $c_t$, and it changes with
$\theta$.
\begin{figure}[!th]
{\par\centering\resizebox*{16cm}{!}{\includegraphics{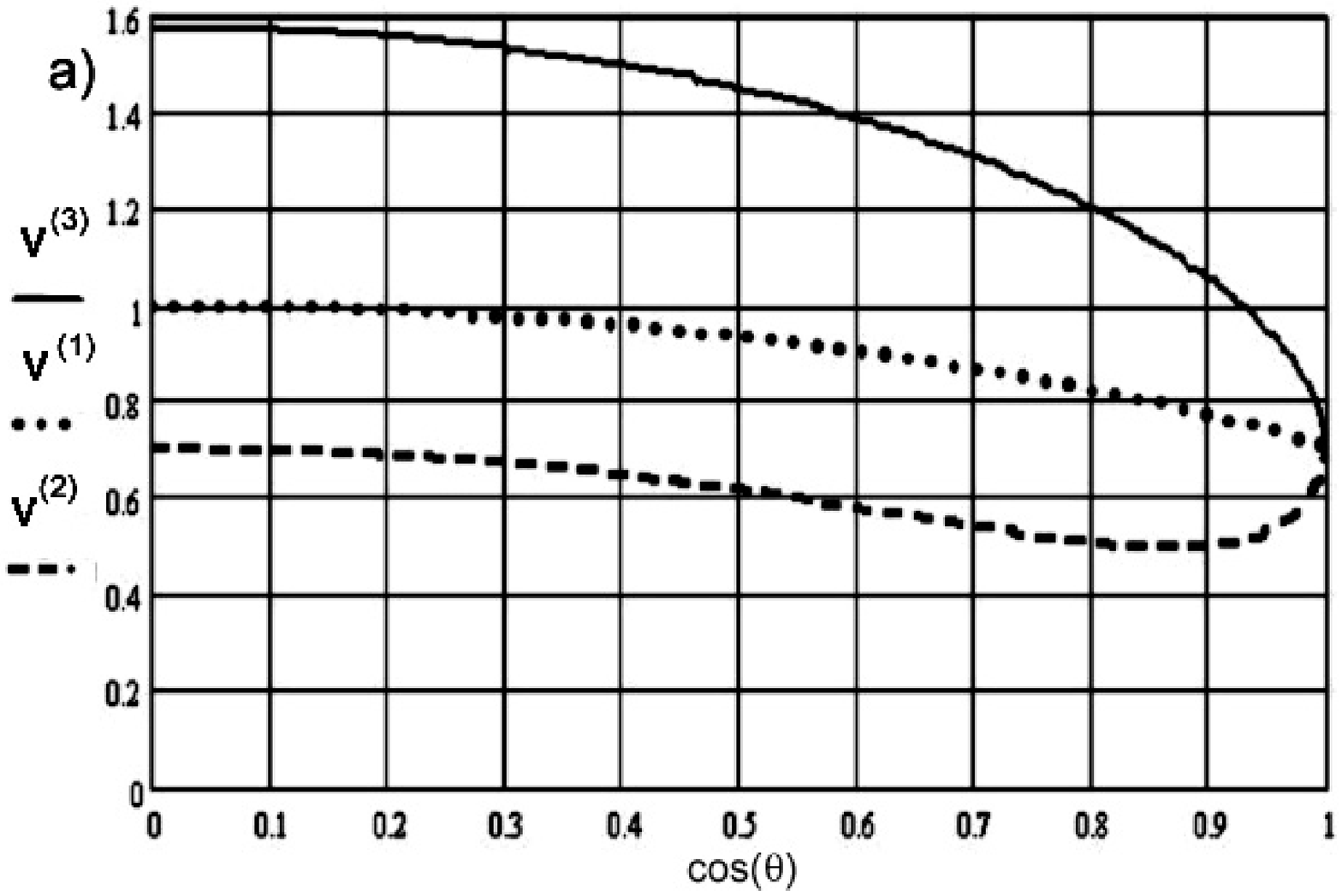}\hspace{15mm}
\includegraphics{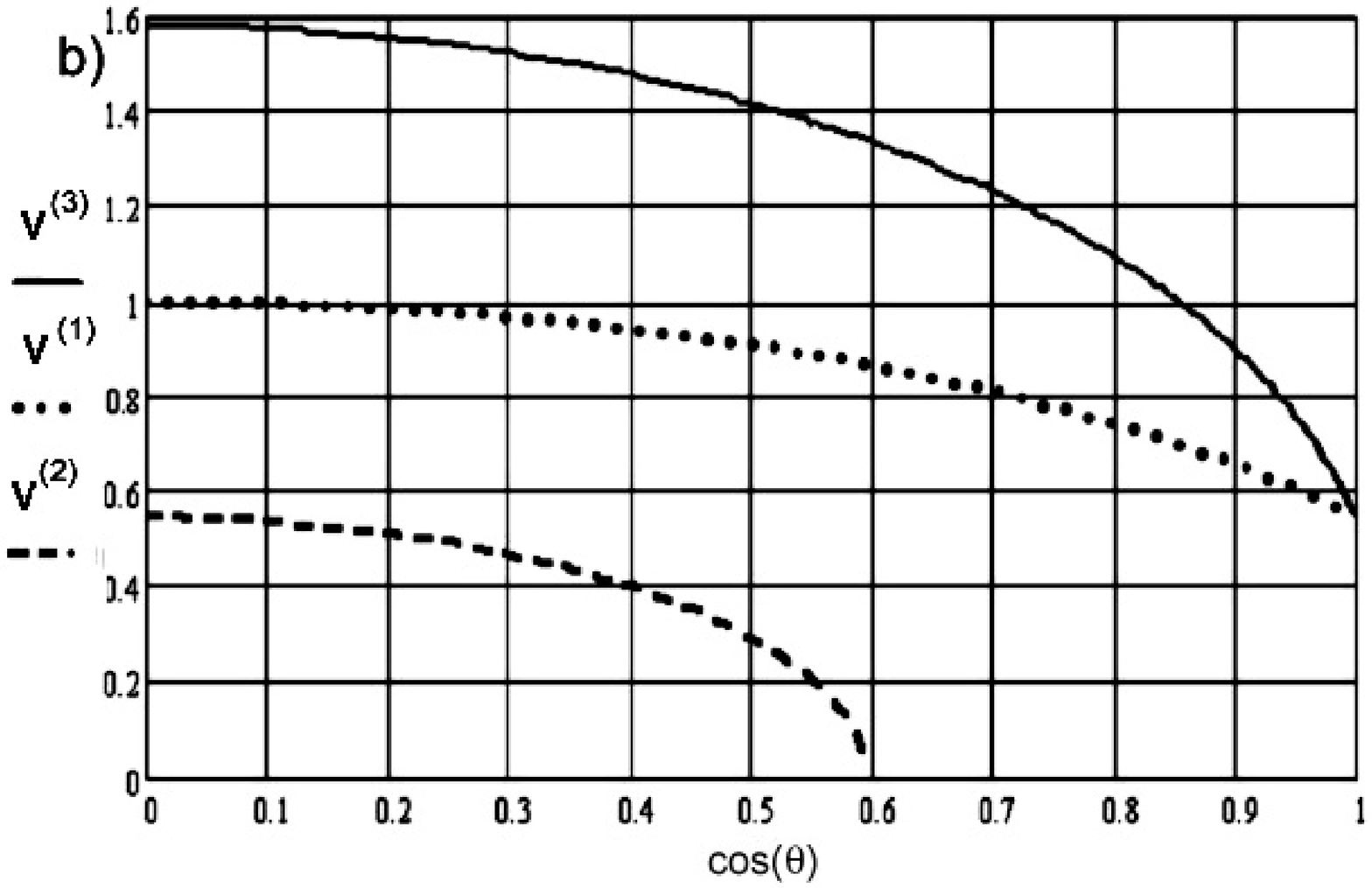}}\par}
\caption{\label{f2} The transverse, $V^{(1)}$, quasi transverse,
$V^{(2)}$, and quasi longitudinal, $V^{(3)}$, speeds of elastic
waves in an anisotropic media in dependence on $\cos\theta$, where
$\theta$ is the angle between wave vector $\kk$ and anisotropy
vector $\av$, for $E=1.5$ and two different anisotropy parameters
a) $\xi=0.5$; b) $\xi=0.7$. The unity on ordinate axis corresponds
to $c_t=\sqrt{\mu/\rho}$.}
\end{figure}

After multiplication of (\ref{eq7}) by $\e^{(2)}$ and $\ka$ we
obtain a system of two equations
\begin{equation}
\label{eq7d} (\Omega^2+ \xi)\alpha^{(2)}
+2\xi(\av\cdot\e^{(2)})(\ka\cdot\av)\beta=0,
\end{equation}
\begin{equation}
\label{eq7e} [\Omega^2 -E+
4\xi(\ka\cdot\av)^2]\beta+2\xi(\av\cdot\e^{(2)})(\ka\cdot\av)\alpha^{(2)}
=0,
\end{equation}
where in \eref{eq7d} we used relation
$(\av\cdot\e^{(2)})^2+(\ka\cdot\av)^2=1$. We see that
polarizations along $\e^{(2)}$ and $\ka$ are not independent. They
combine and create two new hybridized polarizations, which we call
quasi transverse and quasi longitudinal modes and denote by
$\A^{(2,3)}$ like in isotropic case.

The system (\ref{eq7d},\ref{eq7e}) has solutions, if
\begin{equation}\label{eq7f}
[\Omega^2+\xi][\Omega^2-E+
4\xi(\ka\cdot\av)^2]-4\xi^2(\ka\cdot\av)^2(\av\cdot\e^{(2)})^2=0,
\end{equation}
From which it follows
\begin{equation}\label{eq7g}
2(\Omega^{(2,3)2}+\xi)=E+\xi[1-4(\ka\cdot\av)^2]\mp
\sqrt{\{E+\xi[1-4(\ka\cdot\av)^2]\}^2+16\xi^2(\ka\cdot\av)^2(\av\cdot\e^{(2)})^2}.
\end{equation}
Since $(\ka\cdot\av)^2=\cos^2\theta$, and
$(\av\cdot\e^{(2)})^2=\sin^2\theta$ then, because
$\Omega^{(2,3)2}=V^{(2,3)2}/c_t^2-1$, we get that the equation
(\ref{eq7g}) is equivalent to
\begin{equation}\label{eq7h}
V^{(2,3)}=c_t\sqrt{1-\xi+\fr{E+\xi(1-4\cos^2\theta)
\mp\sqrt{[E+\xi(1-4\cos^2\theta)]^2+4\xi^2\sin^2(2\theta)}}{2}}.
\end{equation}

At $\xi\to 0$ their values are
\begin{equation}\label{eq7h3}
V^{(2)}\approx c_t\left(1-\fr\xi2\right)-O(\xi^2),\qquad
V^{(3)}\approx c_l-2\fr{\xi}{c_l} \cos^2\theta+ O(\xi^2),
\end{equation}
where $O(\xi^2)$ denotes a small number proportional to $\xi^2$.
So $V^{(2)}$ can be called quasi transverse and $V^{(3)}$
--- quasi longitudinal speed.

All the speeds $V^{(1)}$, $V^{(2)}$ and $V^{(3)}$ depend on angle
$\theta$. This dependence is shown in Fig.~\ref{f2}. We see that
if the anisotropy parameter $\xi$ is sufficiently large some modes
at small angles $\theta$ cease to propagate, because their speed,
as is shown in Fig.~\ref{f2}b) for quasi transverse mode, does not
exist. This speed becomes imaginary, therefore the wave number of
the mode, $k^{(2)}=\omega/V^{(2)}$, also becomes imaginary, and
the wave does not propagate. Of course it corresponds to too large
anisotropy parameter. Since $E=1.5$ then $\xi=0.7$ means that
anisotropy energy $\zeta$ is larger than the Lam\'e index
$\lambda$, and in some directions the higher is deformation the
less is the elastic energy, which is nonphysical. For smaller
$\xi$ the speed $V_{2}$ is at no angle imaginary.

From (\ref{eq7d}) and (\ref{eq7e}) it follows that polarization of
propagating quasi transverse, $\A^{(2)}$, and quasi longitudinal,
$\A^{(3)}$, modes are
\begin{equation}
\label{eq7d1} \A^{(2)}
=\fr{\xi\sin(2\theta)\e^{(2)}-(\Omega^{(2)2} +
\xi)\ka}{\sqrt{(\Omega^{(2)2} +
\xi)^2+\xi^2\sin^2(2\theta)}},\qquad\A^{(3)} =\fr{(\Omega^{(3)2} +
\xi)\ka-\xi\sin(2\theta)\e^{(2)}}{\sqrt{(\Omega^{(3)2} +
\xi)^2+\xi^2\sin^2(2\theta)}}.
\end{equation}
At small $\xi$ they, as can be expected, are:
\begin{equation}
\label{eq7d2} \A^{(2)} \approx\e^{(2)}+\fr\xi
E\sin(2\theta)\ka,\qquad\A^{(3)} \approx\ka-\fr\xi
E\sin(2\theta)\e^{(2)}.
\end{equation}

\subsection{Reflection from an interface}

Reflection of waves from an interface in anisotropic media is in
general characterized by trirefringency, as was correctly pointed
out in~\cite{leu}. An incident wave at an interface in general
splits into three reflected and three refracted waves, and no wave
is reflected specularly.
\begin{figure}[!t]
{\par\centering\resizebox*{10cm}{!}{\includegraphics{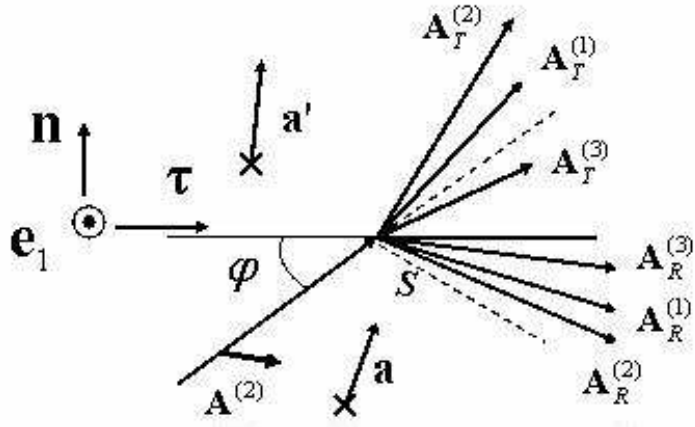}}
\par}
\caption{\label{f4}Splitting of reflected and refracted waves at
an interface between two different anisotropic media when the
incident ia a wave of quasi transverse mode $\A^{(2)}$. S denotes
the specular direction. Anisotropy vector $\av$ has such a
direction that the speed of reflected $\A^{(2)}_R$ mode is higher
than that of the incident one.}
\end{figure}
In Fig. \ref{f4} we present the scheme of reflection and
refraction of a quasi transverse wave from an interface between
two anisotropic media with different anisotropy vectors $\av$ and
$\av'$ and different parameter $\rho$, $\lambda$, $\mu$ and
$\zeta$. The anisotropy vectors in general are not in the
incidence plane. In Fig.~\ref{f4} they are inclined down, so the
reader sees their tails denoted by crosses.

The grazing angles $\varphi^{(i)}$, $\varphi'^{(i)}$ (angles
between wave vectors $\kk^{(i)}_{R,T}$ of reflected and refracted
modes $\A^{(i)}_{R,T}$ and the unit vector $\ta$) in the case when
the incident wave is of mode $j$, are determined from the
relations equivalent to \eref{fi23}:
\begin{equation}\label{fi23f}
\fr{\cos\varphi}{V^{(j)}}=\fr{\cos\varphi^{(i)}}{V^{(i)}_{R}}=\fr{\cos\varphi'^{(i)}}{V^{(i)}_{T}}.
\end{equation}

\begin{figure}[!t]
{\par\centering\resizebox*{16cm}{!}{\includegraphics{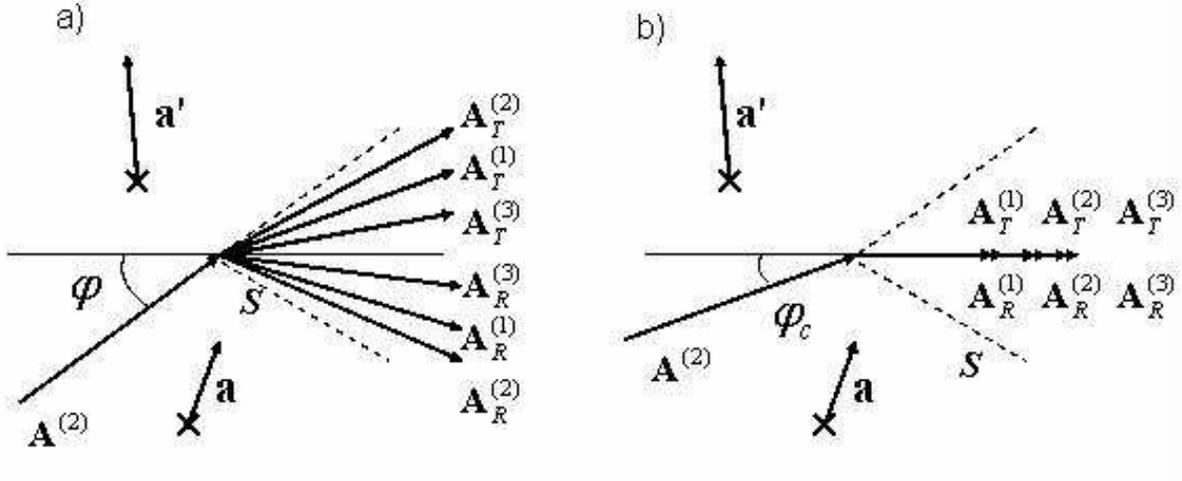}}
\par}
\caption{\label{f5}Reflection, refraction and splitting of waves
at an interface between two different anisotropic media, when the
speeds of refracted waves are higher than that of the incident
quasi transverse wave. a) The grazing angle of the incident wave
is sufficiently high, and all the created waves are able to
propagate in $z$ direction. b) Unreal situation, when the incident
quasi transverse wave is transformed into the surface wave
containing all the three modes.}
\end{figure}

The value of the speed of a wave depends on the angle $\theta$
between the direction of its propagation and the anisotropy vector
$\av$. It may happen that after reflection all the speeds are
higher that the speed $V^{(j)}$ of the incident wave. Then the
grazing angles of all the waves become less that that of the
incident one
 as is shown in Fig. \ref{f5}a), and we can expect that at some
critical angle $\varphi=\varphi_{c}$ all the reflected and
transmitted waves will accumulate into a single surface wave as is
shown in Fig.~\ref{f5}b).

Physically such a result is unacceptable, because the incident
plane wave gives the energy flux toward the interface, therefore
the energy must accumulate in the surface wave and the surface
wave amplitude should increase with the time exponentially. We are
dealing with stationary waves, therefore exponentially growing
functions are excluded from our solutions.

\begin{figure}[!t]
{\par\centering\resizebox*{16cm}{!}{\includegraphics{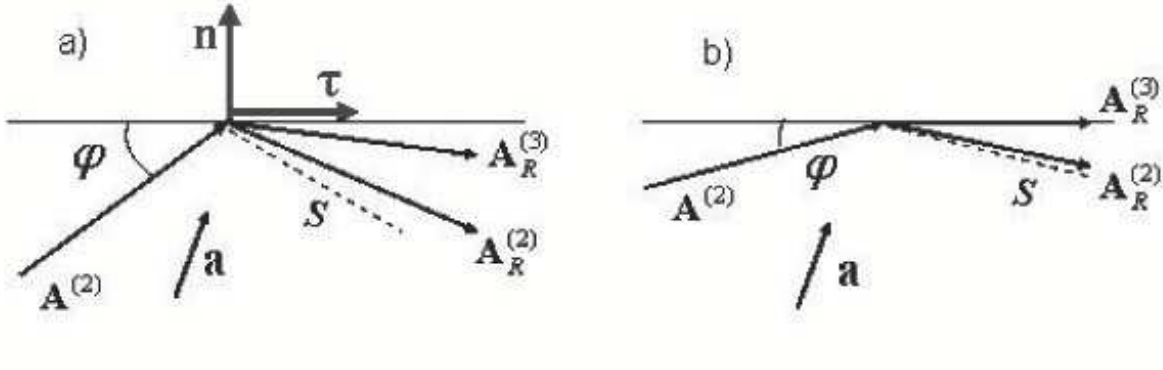}}
\par}
\caption{\label{f6}Reflection of a quasi transverse wave from a
free surface, when anisotropy vector lies in the incidence plane,
and the reflected wave speed is higher than that of the incident
one. Reflection is accompanied with creation of quasi longitudinal
wave. a) The grazing angle of the incident wave is sufficiently
large and both reflected waves can propagate in $z$ direction. b)
The grazing angle of the incident wave is sufficiently small, and
quasi longitudinal wave propagates only along the surface.}
\end{figure}

We should look what is wrong in our logic, considering an example,
in which everything can be solved analytically. The analytical
solution can be found in the case of reflection of a quasi
transverse wave from a free surface, when anisotropy vector lies
in the incidence plane, as is shown in Fig.~\ref{f6}. In this case
we have only two reflected modes: quasi transverse and quasi
longitudinal ones, and to find their reflection amplitudes we need
to solve only system of two linear equations.

\subsection{Reflection of quasi transverse wave from a free surface, when anisotropy vector is in
the incidence plane}

Let's consider reflection of a plane wave of quasi transverse mode
$\A^{(2)}$ from a free surface, when the anisotropy vector has
such a direction, that the reflected quasi transverse wave has
higher speed than the incident one.

The angles of reflected waves are determined by \eref{fi23}
\begin{equation}\label{fi23i}
\fr{\cos\varphi}{V^{(2)}(\theta)}=\fr{\cos\varphi^{(2)}}{V^{(2)}_{R}(\theta^{(2)})}=
\fr{\cos\varphi^{(3)}}{V^{(3)}_{R}(\theta^{(3)})},
\end{equation}
where $\theta$ and $\theta^{(2,3)}$ are the angles between $\av$
and directions of propagation $\ka$ of the incident and
$\ka^{(2,3)}$ of the reflected waves respectively. In these
equations we do not know $V^{(i)}_{R}(\theta^{(i)})$, therefore we
cannot directly find $\varphi^{(i)}$. Instead we have to use these
equations to find both $\varphi^{(i)}$ and $V_{ir}(\theta^{(i)})$
simultaneously.

Let's denote $\av=\ta\cos\varphi_a+\n\sin\varphi_a$,
$\ka=\ta\cos\varphi+\n\sin\varphi$ and
$\ka^{(i)}=\ta\cos\varphi^{(i)}-\n\sin\varphi^{(i)}$ then
$\cos\theta=\av\cdot\ka$ and $\cos\theta^{(i)}=\av\cdot\ka^{(i)}$.
Substitution of $\cos\theta$ and $\cos\theta^{(i)}$ into
\eref{eq7h} and after that into \eref{fi23i} gives transcendent
equations that can be solved numerically.

\begin{figure}[!t]
{\par\centering\resizebox*{10cm}{!}{\includegraphics{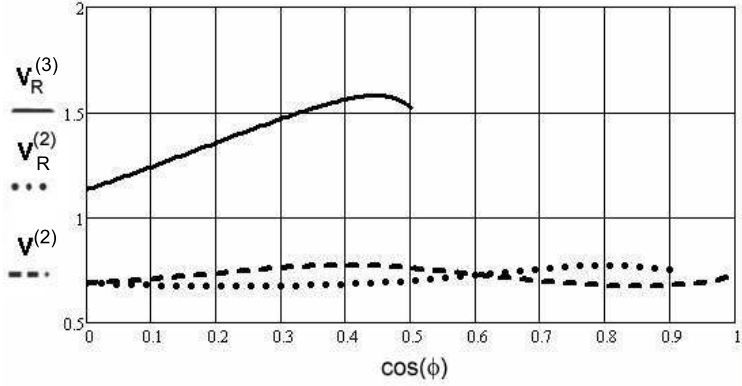}}
\par}
\caption{\label{spr2}Dependence of speeds of the quasi transverse
incident, $V^{(2)}$, quasi transverse reflected, $V^{(2)}_R$ and
quasi longitudinal reflected $V^{(3)}_R$ speeds on $\cos\varphi$
of the grazing incidence angle, when $(\av\cdot\ta)=0.4$;
$\xi=0.4$ and $E=1.5$. We see that in some range of $\cos\varphi$
both reflected speeds are higher than that of the incident one.}
\end{figure}

\begin{figure}[!b]
{\par\centering\resizebox*{8cm}{!}{\includegraphics{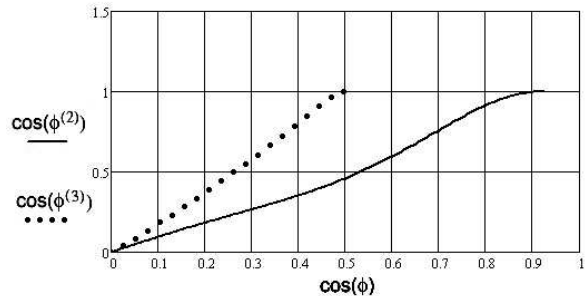}}
\par}
\caption{\label{ang}Dependence of $\cos\varphi^{(2)}$ and
$\cos\varphi^{(3)}$ on $\cos\varphi$. We see that both
$\cos\varphi^{(2)}$ and $\cos\varphi^{(3)}$ reach unity at
$\cos\varphi<1$. It looks as if both waves become of surface type,
when the incident one still remains to be the plain wave.}
\end{figure}

The result of calculations for $\cos\varphi_a=0.4$, $\xi=0.4$ and
$E=1.5$ are shown in Fig-s~\ref{spr2} and \ref{ang}.
Fig.~\ref{spr2} shows dependence of all the speeds on
$\cos\varphi$, and Fig.~\ref{ang} shows dependence of
$\cos\varphi^{(i)}$ on $\cos\varphi$.

We see that at $\cos\varphi>0.92$ no reflected wave can propagate.
What does happen there is the most interesting question!

At $\cos\varphi<0.5$, where both reflected waves do really exist,
we can find their reflection amplitudes. For that we have to solve
the boundary condition equation
\begin{equation}\label{q}
\B^{(2)}+r^{(22)}\B^{(2)}_{R}+r^{(23)}\B^{(3)}_{R}=0,
\end{equation}
where $\B$ is defined like in \eref{eq12b}:
$$\B=-i\exp(-i\kk\rr)\T\Big(\A\exp(i\kk\rr)\Big)=(\n\cdot\kk)\A+\kk(\n\cdot\A)+(E-1)\n(\A\cdot\kk)-$$
\begin{equation}\label{q1}
-\xi\{\av[(\n\cdot\kk)(\av\cdot\A)+(\av\cdot\kk)(\n\cdot\A)]+(\n\cdot\av)
[(\av\cdot\kk)\A+\kk(\av\cdot\A)]\}.
\end{equation}
Multiplying (\ref{q}) by $\n$ and $\ta$, we obtain 2 equations for
two reflection amplitudes. Let's denote $\beta=(\n\cdot\B^{(2)})$,
$\delta=(\ta\cdot\B^{(2)})$, $\beta_i=(\n\cdot\B^{(i)}_R)$ and
$\delta_i=(\ta\cdot\B^{(i)}_R)$ where i=2,3,
\begin{figure}[!t]
{\par\centering\resizebox*{16cm}{!}{\includegraphics{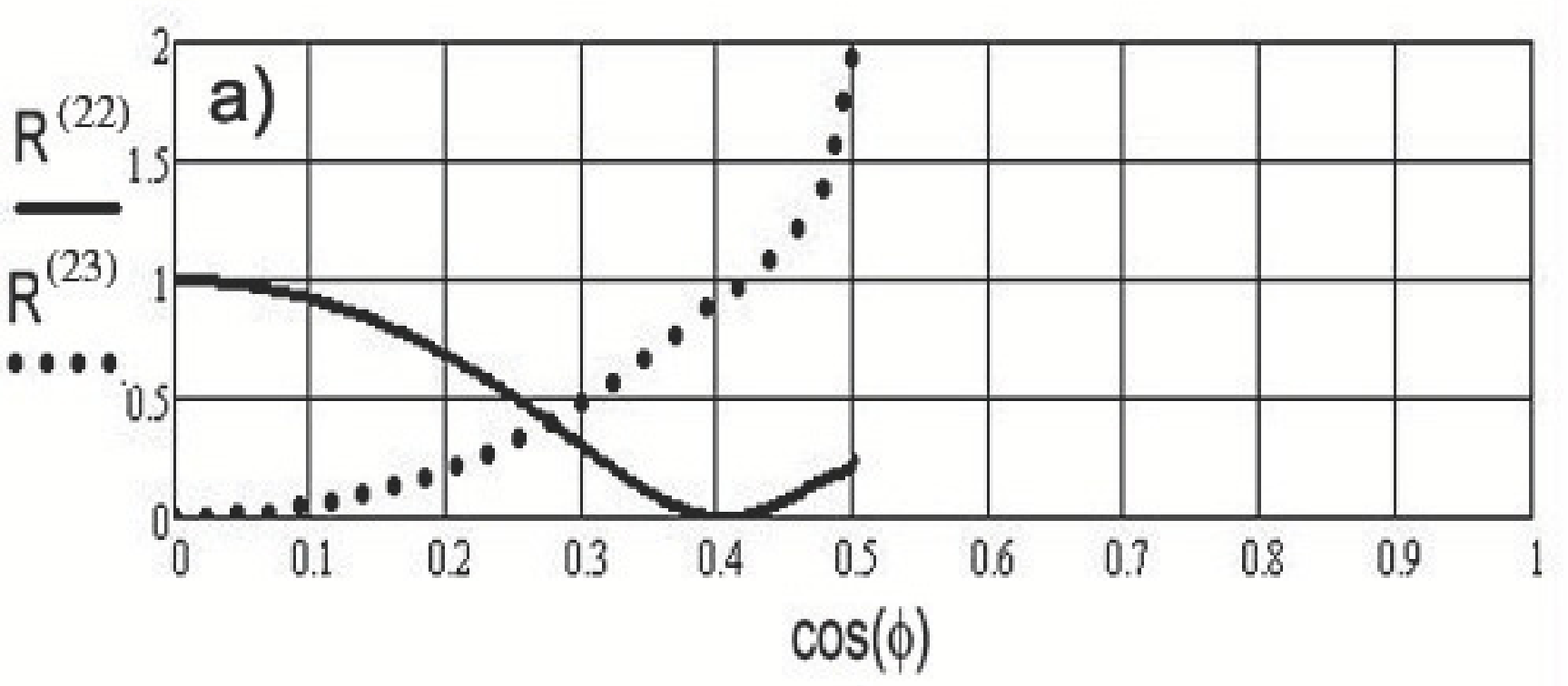}\hspace{15mm}
\includegraphics{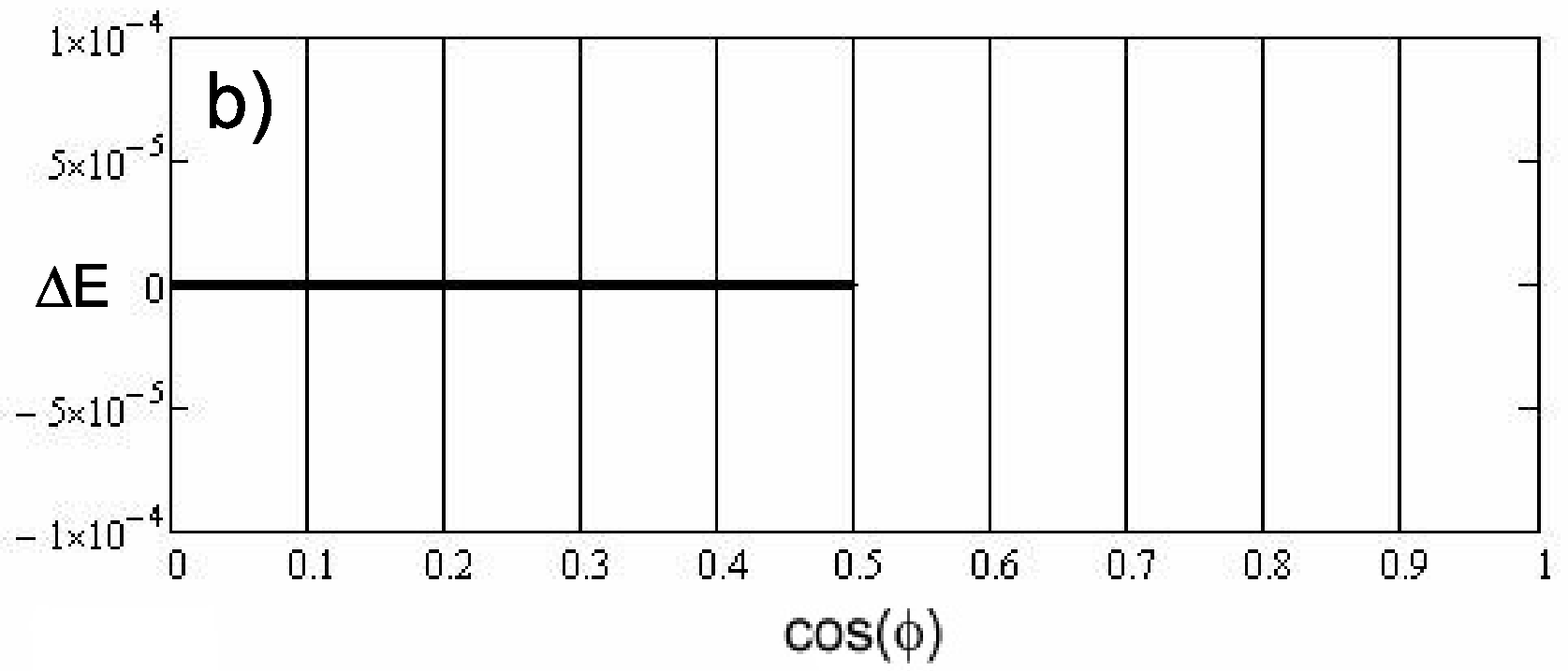}}
\par}
\caption{\label{reng}Dependence of reflectivities
$R^{(22)}=|r^{(22)}|^2$ of quasi transverse and
$R^{(23)}=|r^{(23)}|^2$ of quasi longitudinal waves on
$\cos\varphi$. b) The difference of energies $\Delta E$ of the
incident and reflected waves. We see that all the values are well
calculated only up to $\cos\varphi=0.5$, where longitudinal wave
becomes of the surface type.}
\end{figure}
then Eq.~(\ref{q}) can be represented in matrix form
\begin{equation}\label{qq44}
{\beta\choose\delta}+\lt(\matrix{\beta_{2}&\beta_{3}\cr\delta_{2}&\delta_{3}}\rt){r^{(22)}\choose
r^{(23)}}=0,
\end{equation}
and its solution is elementary. We do not represent the final
analytical result because it does not look sufficiently
informative.

The numerically calculated dependence of reflectivities
$R^{(22)}=|r^{(22)}|^2$ and $R^{(23)}=|r^{(23)}|^2$ on
$\cos\varphi$ is shown in the left panel of Fig.~\ref{reng}. The
correctness of calculations is supported by the panel b), which
demonstrates the energy conservation at reflection. All the
calculations are possible only up to $cos\varphi=0.5$. Above this
value the the wave vector of the quasi longitudinal wave becomes
complex, and equation \eref{fi23i} has no solutions.

\subsection{Waves propagation near a free surface}

To see what happens, above $cos\varphi=0.5$ we have to make
calculations differently. At the interface there are two conserved
values: the frequency $\omega$ and the wave number $k_\|$ along
the interface. It is worth to divide both parts of the Eq.
(\ref{eeq6}) by $\mu k_\|^2$, to introduce the value
\begin{equation}\label{66eq}
\Upsilon=\fr{\omega^2}{c_t^2k_\|^2}=\fr{\omega^2}{k^2c_t^2\cos^2\varphi}=
\fr{V^{(2)2}(\cos\theta)}{c_t^2\cos^2\varphi},
\end{equation}
and the dimensionless wave vector $\tk=\kk/k_\|=\ta+q\n$, where
$q=k_\bot/k_\|$. After that the Eq. (\ref{eeq6}) is transformed to
$$\left[\Upsilon -1-q^2+\xi(\tk\cdot\av)^2\right]\A =
E\tk(\tk\cdot\A)-$$
\begin{equation}
\label{eq66} - \xi\left( \av[(1+q^2)(\av\cdot\A) +
(\tk\cdot\av)(\tk\cdot\A)] + \tk(\tk\cdot\av)(\av\cdot\A)]
\right).
\end{equation}
This equation describes propagation of waves near any, even
fictitious, surface, and it is valid also near the interface. With
it we do not speak about incident and reflected waves. We look for
all possible solutions near the interface and select those which
correspond to our physics. Solution of Eq. (\ref{eeq6}), gave us
wave modes and their speeds, solution of \eref{eq66} will give
wave modes and their $q$ or $k_\bot$. We select in between them,
say, one wave with positive and two waves with negative $q$. They
correspond to the incident and reflected waves. And we find such a
superposition of these waves that satisfies the boundary condition
\eref{q}. Thus we obtain the result absolutely equivalent to that
obtained with Eq. (\ref{eeq6}). In the case of a real $k_\bot$ the
value of $q$ is $\tan\varphi$, but $q$ can be also defined for
arbitrary complex $k_\bot$, and this is the benefit of the
Eq.~\eref{eq66}

All the waves, incident, reflected or surface ones, should satisfy
this equation for the given value of $\Upsilon$, which is
determined by the grazing angle $\varphi$ of the incident wave and
by direction of the vector $\av$. Polarization vector $\A$ is
represented as
\begin{equation}\label{anta}
\A=\alpha\n+\beta\ta.
\end{equation}
To find $\alpha$ and $\beta$ we multiply both parts of Eq.
(\ref{eq66}) by $\n$ and $\ta$, and obtain a system of two linear
homogeneous equations, which has solution, when its determinant is
equal to zero. The resulting equation is a polynomial of the 4-th
order in powers of $q$, and it has 4 roots.

For instance, calculations for $\cos\varphi=0.3$, which is below
0.5, give all the roots $q$ to be real. Two of them are positive:
$q_1=3.18$, $q_4=2.47$; and the other two are negative:
$q_2=-3.6$, $q_3=-1.4$ (such numerations of the roots is for
further convenience). These roots determine all the possible waves
near the surface. The positive roots correspond to waves incident
on the surface, and the negative ones correspond to waves going
away from the surface. The root $q_1$ corresponds to the given
$\cos\varphi=0.3$ of the incident quasi transverse wave
$\A^{(2)}$. The root $q_4$ shows that, if the incident wave were
quasi longitudinal one, its grazing angle would be
$\cos\varphi=0.37$. The negative roots are related to the
reflected waves: $q_2$ to the quasi transverse, $\A^{(2)}_R$, and
$q_3$ --- to the quasi longitudinal, $\A^{(3)}_R$, ones.

When $\cos\varphi=0.6>0.5$ the two roots, $q_3$ and $q_4$, related
to quasi longitudinal waves become complex conjugate:
$q_{3,4}=0.32\mp0.77i$. We have to take into consideration only
$q_3$, which is related to quasi longitudinal surface wave. This
root has correct sign of the imaginary part, which warrants an
exponential decay of the wave away from the interface in the half
space $z<0$. However it does contain also a real part, which seems
to make this root unacceptable. In particle physics the wave
function $\psi\propto\exp(iq'z+q''z)$ at $z<0$ means that there is
a flux of particles $j\propto q'\exp(2q''z)$ toward the surface,
which increases exponentially from $z=-\infty$, and shows that
during propagation from $z=-\infty$ toward $z=0$ the particles are
created from nothing. Intuitively we expect the same of elastic
waves. However elastic waves behave differently, and because of
that we call their properties counter intuitive. The numerical
calculations of the energy flux according to \eref{j1nn} show that
the surface quasi longitudinal wave does not create energy flux
notwithstanding that its $k^{(3)}_\bot$ has a complex value.

The most terrible situation seems to occur after
$\cos\varphi=0.92$. (It is not a fundamental constant. It depends
on direction of the anisotropy vector $\av$ and on values of
parameters $\xi$ and $E$). At some critical
$\varphi_{c1}\approx0.921352$ the value of $q_2$ and therefore of
$k^{(2)}_\bot$ become zero, i.e. our anxieties came true! The
incident plane wave turns into a surface one! However above this
critical point the quasi transverse mode does not become of a
surface type. Its $q_2$ and therefore $k^{(2)}_\bot$ do not
acquire a negative imaginary part. Instead $k^{(2)}_\bot$ remains
real but changes its sign!

Intuitively we can expect that after reverse of the sign of $q_2$
the wave becomes propagating toward the surface. Such a wave
should carry the energy also toward the surface. Nothing like
that! We found that the energy flux of this mode did not change
its sign. Reflected energy flux related to this mode remains
completely equal to the incident flux and opposite in direction.
It can be understood because the energy flux depends not solely on
the wave vector $\kk$ but also on polarization (or oscillation)
direction $\A$ and anisotropy vector $\av$. The direct
calculations show that we have no reason to worry nor about energy
conservation, nor about boundary conditions. They both are
satisfied at $\cos\varphi>\cos\varphi_{c1}$.

However it is not the end of the story. When we decrease $\varphi$
below $\varphi_{c1}$, the value of $q_1=\tan\varphi$ decreases and
the energy flux of the incident wave decreases too. This is
natural. Reflected flux decreases in the same way, though $q_2>0$
steadily increases. But there is a second critical point
$\varphi_{c2}$, where $q_1=q_2$, and the energy flux density of
the incident wave becomes zero! After this point the roles of the
two roots $q_1$ and $q_2$ do exchange. The incident wave gives the
flux away from the surface, and the reflected wave --- toward it.
Of course it means that the incident wave of the mode $\A^{(2)}$
does not exist below $\varphi_{c2}$! All that leads us to an
interesting conclusion, but before going to it let's discuss the
surface waves on a free surface in an anisotropic media.

\subsection{Surface waves}

The first question is: whether the surface waves do exist? From
the very beginning it was found that if we require that a surface
wave to decay away from the surface with a real exponent, the
equation for the speed of the surface wave leads to a complex
value of $c_R$, which means that the surface waves are leaky, and
therefore cannot be accepted as a stationary solution of the wave
equation. However an experience with quasi longitudinal surface
waves had shown that we can accept a complex exponent. Then we may
expect to find a real value for $c_R$.

A surface wave (the Rayleigh one) satisfies the same equation
\eref{eq66} as any other wave, but with $\Upsilon=(c_R/c_t)^2$,
which we earlier (see Eq.~\eref{e14}) denoted  as $x^2$. With it
we rewrite Eq.~\eref{eq66} as
$$\left[x^2-1-q^2+\xi(\tk\cdot\av)^2\right]\A =
E\tk(\tk\cdot\A)-$$
\begin{equation}
\label{6eq66} - \xi\left( \av[(1+q^2)(\av\cdot\A) +
(\tk\cdot\av)(\tk\cdot\A)] + \tk(\tk\cdot\av)(\av\cdot\A)]
\right).
\end{equation}
The easiest way is to find $x$ by try and error method. We suggest
some value of $x=x_1<1$, seek the solution of \eref{6eq66} in the
form \eref{anta}. Multiply both parts by $\n$ and $\ta$, obtain
two homogeneous linear equations for $\alpha$ and $\beta$, find
its determinant, which is a polynomial of 4-th order in powers of
$q$: $D_4(q)$, and find its roots $q_i$ (i=1-4). If all the roots
are complex, we choose two of them with negative imaginary parts,
say $q_2$ and $q_3$. For them we find $\alpha^{(2,3)}$,
$\beta^{(2,3)}$ and $\A^{(2,3)}(q_{2,3})$. After that we use
\eref{eq12b} and obtain $\B^{(2,3)}$. With these vectors we
construct a linear combination, which satisfies boundary
conditions
\begin{equation}\label{qq88}
\gamma\B^{(2)}(q_2)+\delta\B^{(3)}(q_3)=0.
\end{equation}
Multiplication of this equation by $\n$ and $\ta$ gives us again a
system of two equations. It is resolvable, if its determinant
$D(x_1,q_2,q_3)$ is equal to zero.

For an arbitrary chosen $x_1$ the determinant
$D(x_1,q_2,q_3)\ne0$. Instead it is a complex number, say
$D(x_1,q_2,q_3)=y_1+iz_1$. Then we try another $x_2$ till we find
$D(x_2,q_2,q_3)=-y_2-iz_2$, where signs of $y_{1,2}$ and
respectively of $z_{1,2}$ are the same. After that by narrowing
the interval $x_1,x_2$ we find the limiting point $x_0$, where
$D(x_0,q_2,q_3)=0$. The Rayleigh speed is $c_R=x_0c_t$. In the
case of $E=1.5$, $\xi=0.4$ and $\cos\theta_a=0.4$ we got
$c_R=0.6066c_t$.

\section{Conclusion}

We formulated the theory of elastic waves in isotropic media with
the help of complex vector wave functions like in particle
physics. We considered reflection and refraction of waves at an
interface with mode conversion or in other words with double
splitting of the reflected and refracted waves. We had shown that
in the case of a transverse incident wave there is a critical
grazing angle $\varphi_c$, below which the longitudinal reflected
wave becomes of the surface type with a speed in the interval
$(c_t,c_l)$. The speed of the Rayleigh wave is a root of the
equation $1/r=0$ where $r$ is one of reflection amplitudes.

The theory for isotropic media was generalized to anisotropic ones
with a single vector of anisotropy and a specific term in the free
energy of deformation. In such media the transverse and
longitudinal waves become  hybridization, reflection and
refraction at an interface is accompanied in general by triple
splitting of reflected and refracted waves, and all the reflected
waves are nonspecular.

In some cases, when speeds of all the reflected waves are higher
than that of the incident one, a plane wave at some critical
grazing angle $\varphi_{c1}$ can be expected to completely
transform into a surface one, which violates the energy
conservation law. Because of some counter intuitive properties of
elastic waves in anisotropic media such a transformation does not
take place. However there are two critical points in the grazing
angle of the incident wave, which can be considered as a hint that
a nonlinearity should come into play near these points. If the
nonlinearity is included, then the phenomenon, like that one shown
in fig.~\ref{f5}b), could be possible. Transformation of a plane
wave into a surface one should lead to an exponential grows of the
surface wave amplitude, which can be related to such natural
phenomena as the devastating earth quakes.

In many other aspects the wave theory for anisotropic media is a
alike to those for isotropic ones. It predicts the Rayleigh
surface wave on a free surface and the Stonley wave on an
interface. Theses surface waves have complex normal components of
the wave vector, however it does not lead to violation of energy
conservation, because the real part of this normal component does
not create an energy flux from the surface. We would like to
stress that the surface waves, which exponentially decay away from
the interface and at the same time oscillate, are not so called
``leaky surface waves'', because their energy leaks nowhere. The
leaky surface waves cannot exist as a stationary solution of the
wave equation without introduction of some losses because of
nonlinearity or scattering, otherwise they violate the law of
energy conservation.

\section*{Acknowledgement} We are grateful to A.N.Nikitin and
T.I.Ivankina for their interest, and one of us (V.K.I.), is also
grateful to Yu.Kopatch, Yu.Nikitenko, P.Sedyshev and V.Shvetsov
for support.

\end{document}